# The Reification of Patterns in the Design of Description-Driven Systems

J.-M. Le Goff, Z. Kovacs

*CERN, Geneva, Switzerland*

R. McClatchey, F Estrella

*Centre for Complex Cooperative Systems, Univ. West of England, Frenchay, Bristol BS16 1QY UK*

### Abstract

To address the issues of reusability and evolvability in designing self-describing systems, this paper proposes a pattern-based, object-oriented, description-driven system architecture. The proposed architecture embodies four pillars - first, the adoption of a multi-layered meta-modeling architecture and reflective meta-level architecture, second, the identification of four data modeling relationships that must be made explicit such that they can be examined and modified dynamically, third, the identification of five design patterns which have emerged from practice and have proved essential in providing reusable building blocks for data management, and fourth, the encoding of the structural properties of the five design patterns by means of one pattern, the Graph pattern. The CRISTAL research project served as the basis onto which the pattern-based meta-object approach has been applied. The proposed architecture allows the realization of reusability and adaptability, and is fundamental in the specification of self-describing data management components.

# 1. Reflection in Description-Driven Systems

The promotion of implicit system descriptions to become explicit objects is referred to as reification. Reification means to take an abstract concept and regard it as a concrete entity [1]. System descriptions that are represented as objects, can be treated and manipulated as objects. The advantage of reifying system descriptions as objects is that operations can be carried on them, like composing and editing, storing and retrieving, organizing and reading. Since these meta-objects represent system descriptions, their manipulation can result in change in system behavior. As such, reified system descriptions are mechanisms which lead to dynamically modifiable systems.

For reifying language descriptions like class, attribute and association, which themselves act as classes, what is needed is a mechanism for defining the class of a class. In OO programming, the class of a class object is a meta-class. Meta-objects, therefore, are implemented as meta-classes. Object models used in most class-based programming language are fixed and closed. These object models do not allow the introduction and extension of modeling primitives to cater for specific application needs. The concept of meta-classes is a key design technique in improving the reusability and extensibility of these languages.

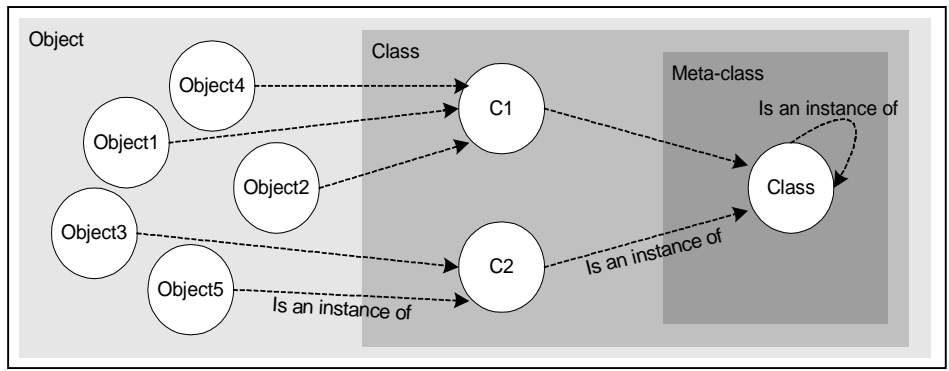

**Figure 1: Graph of Objects, Classes and Meta-classes**

VODAK [2], Prometheus [3], ADAM [4] and OM [5] are some of the next generation DBMSs which have adopted the meta-class approach for tailoring the data model to adapt to evolving specifications. A meta-class may, typically, define properties about object creation, encapsulation, inheritance rules, message passing and the like. A sample meta-class environment (taken from [6]) is shown in Figure 1. In such an environment, every object has a class and every class is an object. Thus object Object1 has class C1 as its class. C1 is also an object, and its class is the meta-class Class. This creates a graph of objects, classes and meta-classes.

A Description-Driven System (DDS) [7] architecture is an example of a reflective meta-layer (i.e. meta-level and multi-layered) architecture, sometimes called a self-describing system. It makes use of meta-objects to store domain-specific system descriptions, which control and manage the life cycles of meta-object instances i.e. domain objects. The separation of descriptions from their instances allows them to be specified and managed and to evolve independently and asynchronously. This separation is essential in handling the complexity issues facing many computing applications and allows the realization of inter-operability, reusability and system evolution as it gives a clear boundary between the application's basic functionalities from its representations and controls. As objects, reified system descriptions of DDSs can be organized into libraries or frameworks dedicated to the modeling of languages in general, and to customizing its use for specific domains in particular.

Reflective techniques have been used by [8] to open up the CORBA architecture. Although the merits of the CORBA architecture have been reaped by many quarters, it is still restrictive and inflexible to future evolutions as its implementation is fixed. The OpenCorba is a reflective open broker which allows users to configure and change the implementation and execution policies of the software bus to cater for their own specific needs. This reflective open broker makes use of meta-classes to reify CORBA semantics and a Meta-Object Protocol (MOP) to allow the dynamic change of run-time behavior. As such, OpenCorba is a clear manifestation of the power of reflection in computing. The OpenCorba initiative is in line with the design of the next generation of middleware, so-called reflective middleware[9]. These middleware are expected to be more re-configurable, and reflection is recognized as the principled approach to provide such re-configurability.

This paper describes an investigation of reified design patterns carried out in the context of the CRISTAL project at CERN, Geneva. This project is not described in detail here but its purpose is to track the workflow processes (and their outcomes) during the highly distributed construction of complex high-energy physics detectors over extended timescales (see [10], [11], [12]) The next section establishes how semantic relationships in description-driven systems can be reified using a set of meta-objects that cater for Aggregation, Generalization, Description,

Dependency and Relationships. In section 3 of this paper the reification of the Graph Pattern is discussed and section 4 investigates the use of this pattern in a three-layer reflective architecture.

## 2. Reifying Semantic Relationships

Objects and classes in OO systems are semantically related. The relationships represent the physical or conceptual connections among the objects involved. Unfortunately, many class-based programming languages have weak or insufficient support for relationships and their inherent semantics [13] & [14]. Most of these languages treat relationships, typically called associations, as second-class objects and let the programmers take the responsibility of implementing them manually. Most often, associations are represented as object pointers. Object pointers are ad-hoc structures without any meaning attached to them. Moreover, the use of object pointers to represent associations distributes the information about the relations among the objects being related, rather than having a single object which can be manipulated as a unit. Thus, object pointers are insufficient in modeling a relationship.

In response to the need to treat associations on an equal footing with classes, a number of published papers have suggested the promotion of the relationship construct as a first-class object. A first-class object is an object which can be created at run-time, can be passed as an actual parameter to methods, can be returned as a result of a function and can be stored in a variable [15]. Reification is used in this paper to promote associations to the same level as classes, thus giving them the same status and features as classes. Consequently, associations become fully-fledged objects with their own attributes to represent their states, and their own methods to alter their behavior.

Different types of relationships representing the many ways inter-dependent objects are related can be reified. The proper specification of the types of relationships that exist among them is essential in managing the relationships and the propagation of operations to the objects they associate. This greatly improves system design and implementation as the burden for handling dependency behavior emerging from relationships is localized to the relationship object. Instead of providing domain specific solutions to handling domain-related dependencies, the relationship objects handle inter-object communication and domain consistency implicitly and automatically.

The next sections discuss four types of relationships, as shown in Figure 2. The relationship classification is divided into two types - structural relationship and behavioral relationship. A structural relationship is a type of relationship which deals with the structural or static aspects of a domain. The Aggregation and the Generalization relationships are examples of this type. A behavioral relationship, as the name implies, deals with the behavioral or dynamic aspects of a domain. Two types of behavioral relationships are explored in this work - the Describes and Dependency relationships.

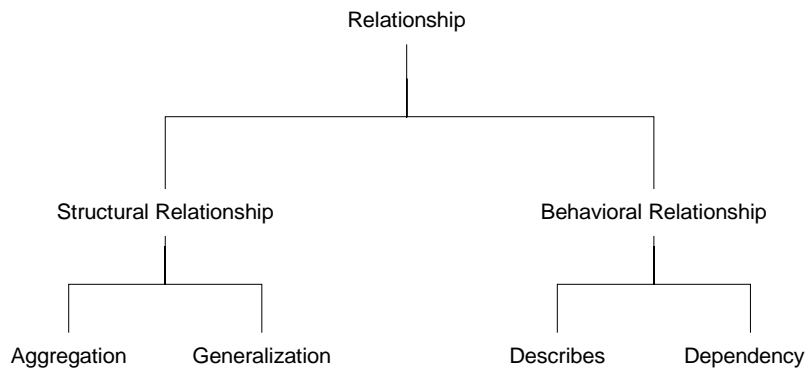

**Figure 2: Relationship classification**

This paper does not provide an exhaustive discussion of all types of relationships, nor does it provide a minimal list of types of associations. Those which are covered are the links which have proved essential and have emerged from a set of five design patterns: the Type Object Pattern, the Tree Pattern, the Graph Pattern, the Publisher-Subscriber Pattern and the Mediator Pattern [16]. Refer to [17], [18] & [19] for a more complete discussion about the taxonomy of semantic relationships.

## 2.1 The Aggregation Meta-Object

Aggregation is a structural relationship between an object whole using other objects as its parts. The most common example of this type of relationship is the bill-of-materials or parts explosion tree [20], representing part-whole hierarchies of objects. The familiar Tree Pattern [13] models the Aggregation relationship and the objects it relates. Aggregated objects are very common, and application developers keep re-implementing the tree semantics for managing part-whole hierarchies. Reifying the Tree pattern provides developers with the Tree pattern meta-object, consequently giving applications with a reusable construct. An essential requirement in the reification of the Tree pattern is the reification of the Aggregation relationship linking the nodes of the tree. For this, aggregation semantics must first be defined.

Typically, operations applied to whole objects are by default propagated to their aggregates. This is a very powerful mechanism as it allows the implicit handling of the management of interrelated objects by the objects themselves through the manner in which they are linked together. By reifying the Aggregation relationship, the three aggregation properties of transitivity, anti-symmetry and propagation of operations can be made part of the Aggregation meta-object attributes and can be enforced by the Aggregation meta-object methods. Thus, the state of the Aggregation relationship and the operations related to maintaining the links among the objects it aggregates are localized to the link itself. Operations like copy, delete and move can now be handled implicitly, automatically and generically by the domain objects irrespective of domain structure.

In addition, the Aggregation meta-object deals with version management among the objects it relates. The versioning strategy can be set (by default) to Propagate, or set to other domain-specific values like No Propagation or Shallow. Setting the version strategy to Propagate means that versioning the whole object automatically versions all its compositions. Setting the version strategy to No Propagation implies that versioning the whole object does not propagate to its compositions. A Shallow version strategy indicates that the whole object and the Aggregation object are versioned, while its compositions remain the same. Another advantage of the Aggregation meta-object is its handling of complexity. Complexity is reduced not only because a group of objects is treated as a single composition object, but also the effect on the root of composition is automatically triggered onto its components without further user action or intervention.

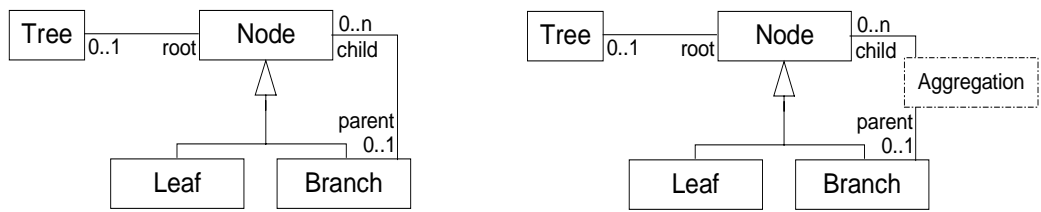

**Figure 3: The Tree Pattern with Reified Aggregation Relationship**

Figure 3 illustrates the inclusion of the reified Aggregation relationship in the Tree pattern. In the diagram, the reified Aggregation relationship is called Aggregation, and is the link between the nodes of the tree. The Aggregation meta-object manages and controls the link between the tree nodes, and enforces the propagation of operations from parent nodes to their children. Consequently, operations applied to branch nodes are by default automatically propagated to their compositions.

## 2.2 The Generalization Meta-Object

Generalization is a structural relationship between a superclass and its subclasses. The semantics of generalization revolve around inheritance, type-checking and reuse, where subclasses inherit the attributes and methods defined by their superclass. The subclasses can alter the inherited features and add their own. This results in a class hierarchy organized according to similarities and differences. The graph of meta-classes, classes and object shown in Figure 1 is an example of a class hierarchy.

Unlike the Aggregation relationship, the generalization semantics are known and implemented by most programming languages, as built-in constructs integrated into the language semantics. This paper advocates extending the programming language semantics by reifying the Generalization relationship as a meta-object. Consequently, programmers can access the generalization relation as an object, giving them the capability of manipulating superclass-subclass pairs at run-time. As a result, application programs can utilize mechanisms for dynamically creating and altering the class hierarchy, which commonly require re-compilation for many languages.

Similar to the Aggregation relationship, generalization exhibits the transitivity property in the implicit propagation of attributes and methods from a superclass to its subclasses. The transitivity property can also be applied to the propagation of versioning between objects related by the Generalization relationship. Figure 4 illustrates the Tree pattern with the Generalization and Aggregation relationships between the tree nodes reified.

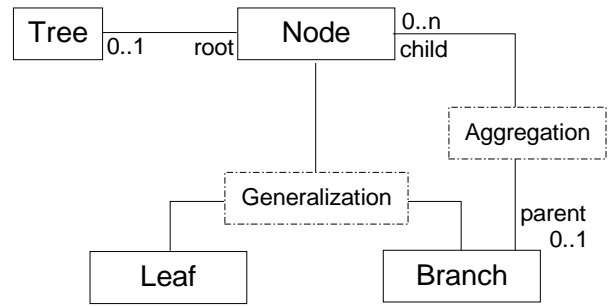

**Figure 4: Reification of the Generalization and Aggregation Relationships**

## 2.3  The Describes Meta-Object

The Type Object pattern of [21] illustrates an important type of relationship. In essence the Type Object pattern has three elements, the object, its type and the Describes relationship which relates the object to its type. The Type Object pattern illustrates the link between meta-data and data and the Describes relationship that relates the two. Consequently, this pattern links layers of multi-level systems. The upper meta-level meta-objects manage the next lower layer's objects. The meta-data that these meta-objects hold describe the data the lower level objects contain. The Type Object pattern is a very useful and powerful tool for run-time specification of domain types.

The reification of the Describes relationship as a meta-object provides a mechanism for explicitly linking object types to objects. This strategy is similar to the approach taken for the Aggregation and Generalization relationships. The Describes meta-object provides developers with an explicit tool to dynamically create and alter domain types, and to modify domain behavior through run-time type-object alteration.

Unlike the Aggregation relationship and the Generalization relationship, the Describes relationship does not exhibit the transitivity property. This implies that the propagation of some operations is not the default behavior as it cannot be inferred for the objects and their types. For example, versioning a type does not necessarily mean that objects of that type need to be versioned as well. In this particular case, it is the domain which decides whether the versioning should be propagated or not. Thus, the Describes meta-object should include a mechanism for specifying propagation behaviour. Consequently, programmers can either accept the default relationship behaviour or override it to implement domain-specific requirements.

Figure 5 illustrates the transformation of the Type Object pattern with the use of a reified Describes relationship. The object pointer (in Figure 5a) is dropped as it is insufficient to represent the semantics of the link relating objects and their types. Instead, the Describes meta-object (in Figure 5b) is used to manage and control the Type Object pattern relationship.

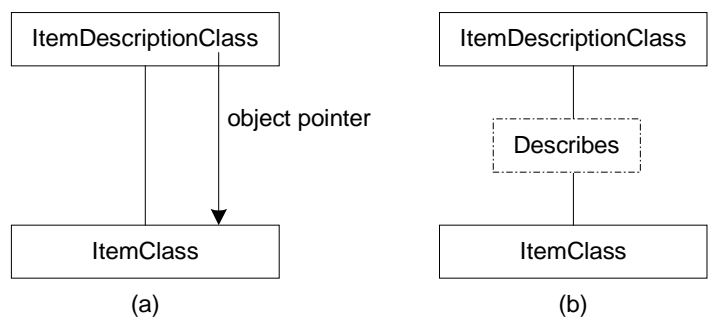

**Figure 5: The Type Object Pattern with Reified Describes Relationship**

## 2.4 The Dependency Meta-Object

The Publisher-Subscriber pattern models the dependency among related objects. Summarizing this pattern, subscribers are automatically informed of any change in the state of its publishers. Thus, the association between the publisher and the subscriber manages and controls the communication and transfer of information between the two.

Most application programs implement mechanisms to deal with domain consistency and propagation of change. Reifying the Publisher-Subscriber dependency association (from hereafter referred to as Dependency association), these mechanisms can be generically implemented and automatically enforced by the Dependency meta-object itself and taken out of the application code. This represents a significant breakthrough in the simplification of application codes and in the promotion of code reuse.

The reification of the Dependency relationship is significant in that it provides an explicit mechanism for handling change management and consistency control of data. The Dependency meta-object can be applied to base objects, to classes and types, to components of distributed systems and even to meta-objects and meta-classes. This leads to an homogeneous mechanism for handling inter-object dependencies within and between layers of multi-layered architectures.

The Event Channel of the Publisher-Subscriber pattern [22] and the Mediator of the Mediator pattern are realizations of the Dependency relationship. The Event Channel, is an intervening object which captures the implicit invocation protocol between publishers and subscribes. The Mediator encapsulates how a set of objects interact by defining a common communication interface. By utilizing the Describes relationship, an explicit mechanism can be used to store and manage inter-object communication protocols. Figure 6 illustrates the use of reified Dependency meta-object in the Publisher-Subscriber pattern (a) and the Mediator pattern (b).

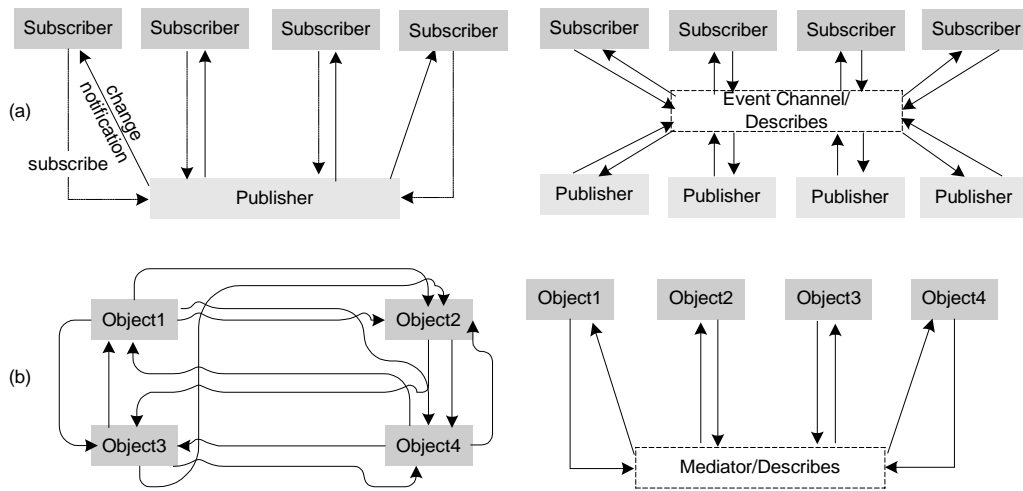

**Figure 6: The Event Channel and the Mediator as Reified Dependency**

## 2.5 The Reified Relationship Meta-Object Protocol

The relationship MOP defines the set of methods for querying information about the relationship and for manipulating the relationship meta-objects [6]. It is the relationship MOP, which controls and manages the relationship meta-object and its object instances. As the relationship semantics involve both the structural and dynamic aspects of a domain, the relationship meta-objects and their MOPs are mechanisms for defining and configuring system behaviour.

Before proceeding to the relationship MOP specification, it is essential that the basic set of relationship attributes be defined first. The relationship is a link between parent nodes and children nodes, as shown in Figure 7.

The figure shows the cardinality of the relationship with respect to its parents and its children. The Aggregation, Generalization, Describes and Dependency relationships are restricted to having one parent node. Having a single parent node for the Aggregation relationship is defined by its semantics. With regards to Generalization, Describes and Dependency links, the cardinality of one for the parent is a design decision which simplifies the relationship model. Allowing many superclasses in the Generalization link results in multiple inheritance, allowing many types in the Describes link creates a multi-typed object, and allowing many publishers in the Dependency link implies that subscribers get data from many subjects. Multiple inheritance, multi-typed object and multiple publisher specifications require further research with regards to compatibility, composability and conflict resolution, and are beyond the scope of this paper.

Reifying relationships as meta-objects is a fundamental step in the reification of design patterns. The four relationship meta-objects discussed manifest the links that exist among the objects participating in the five design patterns listed in the introduction to this section. With the use of reified relationships, these five patterns can be modeled as a single graph, using the Graph pattern. Consequently, the five design patterns are structurally reified as a Graph pattern with the appropriate relationship meta-object to represent the semantics relating the pattern objects.

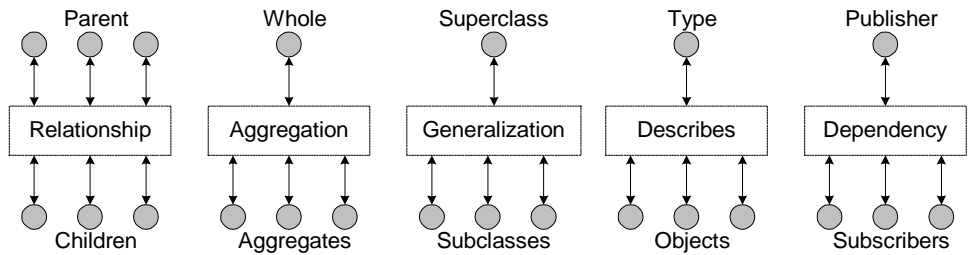

Figure 7: The cardinality of Relationship

## 3. The Reified Graph Pattern

The graph and tree data structures are natural models to represent relationships among objects and classes. As the graph model is a generalization of the tree model, the tree semantics are subsumed by the graph model. Consequently, the graph specification is applicable to tree representations. The compositional organization of objects using the Aggregation relationship forms a graph. Similarly, the class hierarchy using the Generalization relationship creates a graph. These two types of relationships are pervasive in computing, and the use of the Graph pattern to model both semantics provides a reusable solution for managing and controlling data compositions and class hierarchies.

The way dependent objects are organized using the Dependency association also forms a graph. A dependency graph is a representation of how interrelated objects are organized. Dependency graphs are commonly maintained by application programs, and their implementations are often buried in them. The reification of the Dependency meta-object objectifies the dependency graph and creates an explicit Publisher-Subscriber pattern. Consequently, the dependency graph is treated as an object, and can be accessed and manipulated like an object. The same argument applies to the Describes relationship found in the Type Object pattern. The link between objects and their type creates a graph. Reifying the Describes relationship results in the reification of the Type Object pattern. With the reification of the Type Object pattern, the resulting graph object allows the dynamic management of object-type pairs. This capability is essential for environments with unknown or dynamically changing user requirements.

The UML diagram of the Graph meta-object is shown in Figure 8. The Node class represents the entities of the domain - objects, classes, data, meta-data or components. The Relationship is the reification of the link between the Nodes. The aggregated links between the Node and the Relationship are bidirectional. Two roles are defined for the two aggregated associations - that of the parent, and that of the child. A relationship has at most one parent node, and a parent node can have zero or more relationships. From the child nodes' point of view, a relationship can have at least one child, and a node is a child of zero or more relationships. The parent aggregation, symbolized by the shaded diamond, implies that the lifecycle of the relationship is dependent on the lifecycle of the parent node. Analogously, the child aggregation behaves in the same manner.

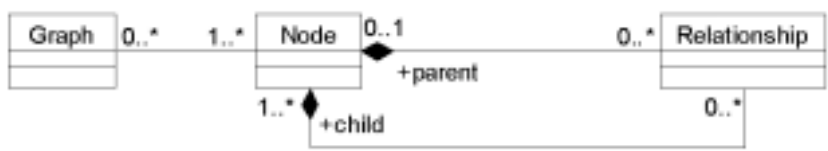

Figure 8: UML Diagram of the Graph Meta-Object

The use of reflection in making the Graph pattern explicit brings a number of advantages. First of all, it provides a reusable solution to data management. The reified Graph meta-object manages static data using Aggregation and Generalization meta-object relationships, and it makes persistent data dependencies using Describes and Dependency relationships. As graph structures are pervasive in many domains, the capture of the graph semantics in a pattern and objectifying them results in a reusable mechanism for system designers and developers. This makes the Graph meta-object a useful guideline applicable to many situations and domains. Another benefit of having a single mechanism to represent compositions and dependencies is its provision for inter-operability. With a single framework sitting on top of the persistent data, clients and components communicate with a single known interface. This greatly simplifies the overall system design and architecture, thus improving system maintainability. Moreover, clients and components can be easily added as long as they comply with the graph interface.

Complexity is likewise managed as related objects are treated singly and uniformly. Firstly, the semantic grouping of related objects brings transparency to clients' code. Secondly, the data structures provided by the Graph meta-object organize data into atomic units which can be manipulated as single objects. Objectifying graph relationships allows the implicit and automatic propagation of operations throughout a single grouping. Another benefit in the use of the reified graph model is its reification of the link between meta-data and data. As a consequence, the Graph meta-object does not only provide a reusable solution for managing domain semantic groupings, but can also be reused to manage and control the links between layers of meta-level architectures.

## 4. Patterns, Relationships and Descriptions

This paper proposes that the reified Graph pattern provides the necessary building block in managing data in any DDS architecture. Figure 9 illustrates the proposed description-driven architecture. The architecture on the left hand side is typical of layered systems such as the multi-layered architecture specification of the OMG [24]. The relationship between the layers is Instance-of. The instance layer contains data which are instances of the domain model in the model layer. Similarly, the model layer is an instance of the meta-model layer. On the right hand side of the diagram is another instance of model abstraction. It shows the increasing abstraction of information from meta-data to model meta-data, where the relationship between the two is also Instance-of. These two architectures provide layering and hierarchy based on abstraction of data and information models.

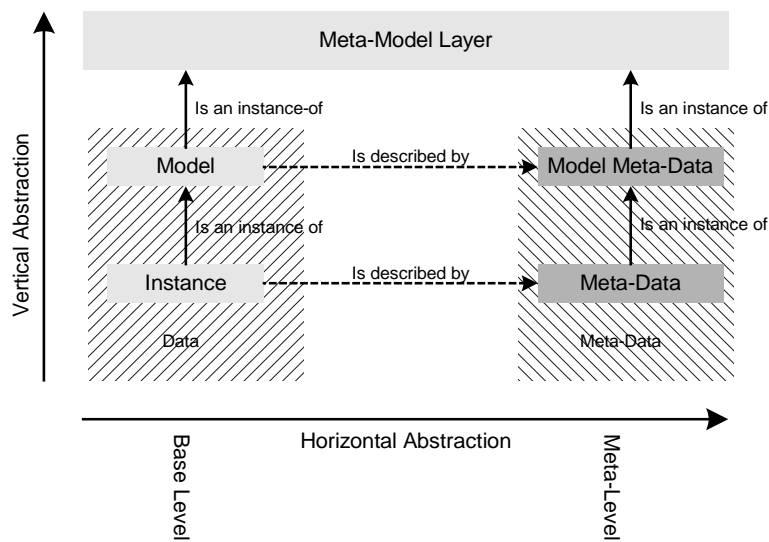

**Figure 9: A Three-layer Reflective DDS Architecture**

This paper proposes an alternative view by associating data and meta-data through description. The Type Object pattern makes this possible. The Type Object pattern is a mechanism for relating data to information describing data. The link between meta-data and data using the Describes relationship promotes the dynamic creation and specification of object types. The same argument applies to the model meta-data and its description of the domain model through the Describes relationship. These two horizontal dependencies result in an horizontal meta-level architecture where the upper meta-level describes the lower level. The combination of a multi-layered architecture based on the Instance-of relationship and that of a meta-level architecture based on the Describes relationship results in what is referred to as a DDS architecture.

The reified Graph pattern provides a reusable mechanism for managing and controlling data compositions and dependencies. The graph model defines how domain models are created. Similarly, the graph model defines how meta-data are instantiated. By reifying the semantic grouping of objects, the Graph meta-object can be reused to hold and manage compositions and dependencies within and between layers of a DDS (see figure 10). The meta-level meta data are organized as a meta-level graph. The base-level data are organized as a base-level graph.

Relating these two graphs forms another graph, with the nodes related by the Describes relationship. These graphs indicate the reuse of the Graph pattern to model relationships in a DDS architecture.

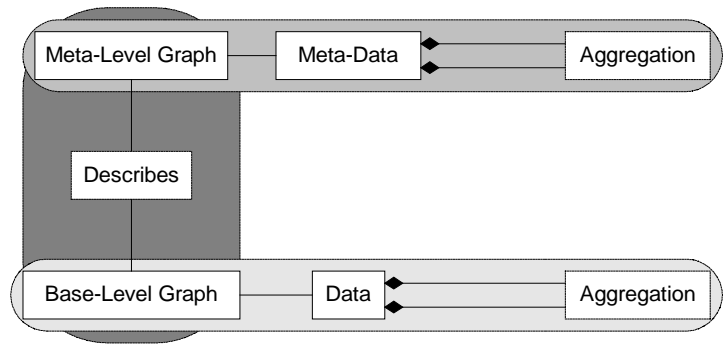

**Figure 10: The Reuse of the Reified Graph Pattern in a DDS**

## 5. Conclusions

As shown in figure 11, he reified Graph pattern and the reified relationships enrich the meta-model layer by giving it the capability of creating and managing groups of related objects [25], [26]. The extension of the meta-model layer to include constructs for specifying domain semantic groupings is the proposition of this paper. The meta-model layer defines concepts used in describing information in lower layers. The core OMG/UML meta-model constructs include Class, Attribute, Association, Operation and Component meta-objects. The inclusion of the Graph meta-object in the meta-model improves and enhances its modeling capability by providing an explicit mechanism for managing compositions and dependencies throughout the architecture. As a result, the reified Graph pattern provides an explicit homogeneous mechanism for specifying and managing data compositions and dependencies in a DDS architecture.

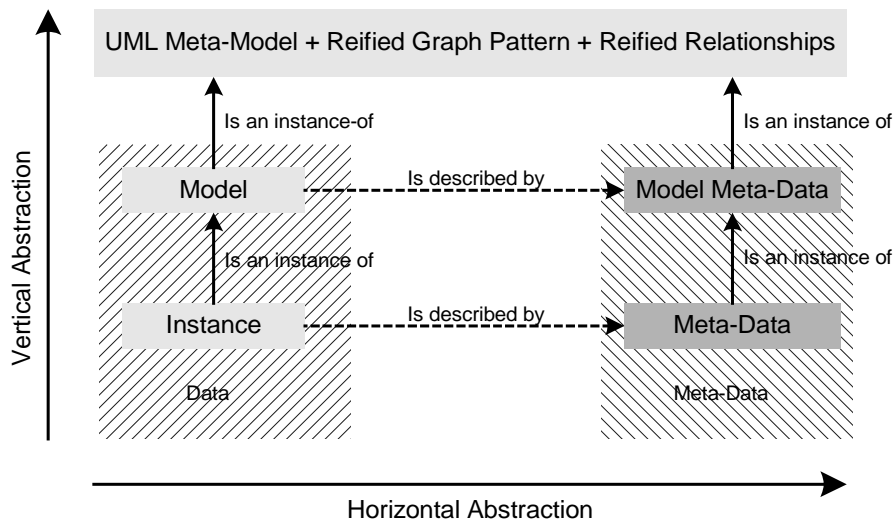

**Figure 11: Extending the UML Meta-Model using a Reified Graph Pattern.**

This paper has shown how reflection can be utilized in reifying design patterns. Reified design patterns provide explicit reusable constructs for managing domain semantic groupings. These pattern meta-objects are then used as building blocks for describing compositions and dependencies in a three-layer reflective architecture - the DDS architecture. The judicious use and application of the concepts of reflection, design patterns and layered models create a dynamically modifiable system which promotes reuse of code and design, which is adaptable to evolving requirements, and which can cope with system complexity. In conclusion, it is interesting to note that the OMG has recently announced the so-called Model Driven Architecture as the basis of future systems integration [27]. Such a philosophy is directly equivalent to that expounded in this and earlier papers on the CRISTAL description-driven architecture.

## Acknowledgments

The authors take this opportunity to acknowledge the support of their home institutes. N. Baker, A. Bazan, P. Brooks, G. Chevenier, T. Le Flour, C. Koch, S. Lieunard, S. Murray, L. Varga, G. Organtini and N. Toth are thanked for their assistance in developing the CRISTAL prototype.